\documentclass[useAMS,usenatbib]{mn2e}
\usepackage{amsmath,amsfonts,graphicx,txfonts,fleqn,color,tabularx,verbatim,fixltx2e}
\newcommand {\B}[1] {\boldsymbol{#1}}

\newcommand  *{\D}[1] {\dot{\B{#1}}}
\newcommand  *{\DD}[1] {\ddot{\B{#1}}}
\renewcommand*{\H}[1] {\hat{\B{#1}}}

\title[Assembling compact planetary systems]{Understanding the assembly of {\it Kepler's} compact planetary systems}
\author[Hands, Alexander \& Dehnen ]{T.O.Hands\thanks{email:
tom.hands@le.ac.uk}, R.D.Alexander and W.Dehnen
\\Department of Physics \& Astronomy, University of Leicester, University Road, Leicester, LE1 7RH, UK}
\begin{document}
\voffset=-0.65in

\pagerange{\pageref{firstpage}--\pageref{lastpage}} \pubyear{2014}

\maketitle

\label{firstpage}

\begin{abstract}
The Kepler mission has recently discovered a number of exoplanetary
systems, such as Kepler-11 and Kepler-32, in which ensembles of several
planets are found in very closely packed orbits (often within a few
percent of an AU of one another). These compact configurations present a challenge for traditional planet formation and migration scenarios.  We present a dynamical study of the assembly of these systems, using an N-body method which incorporates a parametrized model of 
planet migration in a turbulent protoplanetary disc.  We explore a wide 
parameter space, and find that under suitable conditions it is possible to 
form compact, close-packed planetary systems via 
traditional disc-driven migration.  We find that simultaneous migration of multiple planets is a viable mechanism for the assembly of tightly-packed planetary systems, as long as the disc provides significant eccentricity damping and the level of turbulence in the disc is modest. We discuss the implications of our preferred parameters for the protoplanetary discs in which these systems 
formed, and comment on the occurrence and significance of mean-motion resonances in 
our simulations.

\end{abstract}


\section{Introduction}
Of the myriad of recent advances in our study of extra-solar planets, perhaps the most interesting is the discovery of large numbers of systems of multiple planets.  The first multiple-planet system around a main sequence star, Upsilon Andromedae, was reported by \citet{Butler1999}, and we now know that many, if not most, planets form in multiple systems.  In the final {\it Kepler} data release \citep{Burke2014} there were 2738 planet candidates around 2017 unique stars.  475 (24\%) of these stars host multiple candidates, while 1196 (44\%) of the candidates are found in multiple-planet systems.  Statistical considerations suggest that the false positive rate in Kepler's multiple-planet systems is low \citep{Lissauer2012}.  When one considers non-detections, it seems likely that the majority of planets form in multi-planet systems. 

Among this avalanche of new data, arguably the most surprising discovery was a new class of compact, tightly-packed systems.  These systems, for which Kepler-11 is the prototype \citep{Lissauer2011,Lissauer2013}, consist of several Neptune- or super-Earth-size planets, typically within a few tenths of an AU of their host star, and often within a few hundredths of an AU of one another.  They are invariably dynamically cold, with low eccentricities and low mutual inclinations. Mean-motion resonances between adjacent planets are seen in some cases, but not all \citep[e.g.,][]{Lissauer2013,Swift2013}, and many of these compact systems appear to be close to dynamical instability \citep{Deck2012}.  Compact planetary systems therefore represent a striking contradiction: they apparently require delicate assembly in order to avoid being destroyed by dynamical instabilities, but their prevalence suggests a robust formation mechanism.

The extreme architectures of these systems have led a number of authors to consider {\it in situ} formation models \citep[e.g.,][]{Hansen2012,Hansen2013,Chiang2013}, as these naturally avoid the complications and uncertainties inherent in migrating systems of multiple planets. However, \cite{Swift2013} show that the dust sublimation radius for Kepler-32 is outside the innermost planet's orbit for the length of all but the most unrealistic disc lifetimes, suggesting that at least some planets in compact systems could not have formed at their present locations. Moreover, conventional protoplanetary disc models contain too little mass at small radii to form systems such as Kepler-32 {\it in situ} \citep{Swift2013}. \cite{Hansen2013} used Monte Carlo simulations of {\it in situ} formation to show that this mechanism can qualitatively reproduce the distribution of tightly-packed systems -- albeit with a slight shift towards longer periods -- while \cite{Hansen2014} suggested that tidal dissipation my be responsible for bringing planets inside the expected dust sublimation radius to shorter-period orbits. More recently, \cite{Raymond2014} have argued against  {\it in situ} formation, on the basis that the range of disc models required for the formation of Kepler's sample of tightly-packed super-Earths in this way is unrealistic.

The natural alternative to {\it in situ} formation is that hot super-Earths formed at larger radii and migrated inwards. These planets are of sufficiently low mass that they are unlikely to open a gap in a disc, and hence undergo Type I migration \citep{Kley2012}. \cite{Raymond2008} made predictions of observational signatures that might help to distinguish between {\it in situ} and migration models, and find that mean-motion resonances would be expected if such systems formed by migration. Several authors have conducted simulations exploring the scenario in which hot super-Earths form from inwardly migrating planetary embryos. \cite{Terquem2007} performed calculations which can form 2--5 super-Earths in very tight orbits since their migration is halted at the inner edge of the disc, whilst \cite{McNeil2010} found that super-Earths of up to 3--4$M_\oplus$ form readily in a sufficiently massive disc. \cite{Cossou2014} have explored the problem with a more recent model of type-I migration and found that super-Earth sized objects can form from embryos migrating from 1--20AU and either pile up at the inner edge of the disc they are embedded in, or become giant enough to migrate outwards and become the core of a giant planet. Note that the line between {\it in situ} formation and formation further out in the disc can be blurred once the embryos have migrated sufficiently far in, since they may continue to accrete material once the disc has dissipated via collisions with other bodies \citep{Cossou2014}.

In this paper we investigate assembling compact, tightly-packed planetary systems by traditional, disc-driven migration. We consider a scenario in which planets are formed fully further out in the disc and then migrate inwards. We adopt an N-body approach, using a variety of parametrized forces designed to mimic those that each planet would experience in a real protoplanetary disc. We consider the influence of each of the main elements of the Type I regime: planetary migration, eccentricity damping and disc turbulence. This results in a large parameter space with many inherent uncertainties, so we use a statistical approach, running large numbers of models of individual systems in order to understand what balance of parameters is most conducive to building these systems. This approach allows us to capture the essential physics of the simultaneous migration of multiple planets, while avoiding the computational cost of full hydrodynamic calculations. In essence our calculations are a proof-of-concept, designed simply to understand whether simultaneous migration of multiple planets is a viable model for the assembly of tightly-packed planetary systems.

\section{Numerical method}
Following the simultaneous migration of many planets in a hydrodynamic calculation is computationally expensive, and the large parameter space makes this approach infeasible here.  We instead model the assembly of compact multi-planet systems using an N-body approach, employing parametrized forces to mimic the effects of disc-driven migration, eccentricity damping and turbulent forcing. 

\subsection{N-body integrator}
We compute gravitational interactions (both star-planet and planet-planet) using a direct summation N-body code.  We adopt a modified 2nd-order kick-drift-kick leapfrog method (see appendix \ref{integrator}) for time integration, with adaptive time-stepping to ensure numerical accuracy whilst minimising the computational cost\footnote{Strictly, the use of adaptive time-stepping violates the symplectic nature of the integrator but in practice this is unimportant, as the system is in any case non-Hamiltonian due to our use of non-conservative damping forces.}.  The star and planets are modelled as point masses, and no gravitational softening is used.  We assign physical radii to all the particles, but these are used only to identify physical collisions between particles.

\subsection{Migration and eccentricity damping} \label{sec:migration}
Gravitational torques from a protoplanetary disc drive both planet migration and eccentricity damping.  The low planet masses considered here are expected to migrate in the Type I regime \citep{Kley2012} where this torque is given by \citep[e.g.,][]{Tanaka2002,Paardekooper2009}
\begin{equation}\label{eq:type1}
\Gamma = -C \frac{q^2}{h^2}  \Sigma_p  a^4  \Omega_p^4 \, .
\end{equation}
Here $a$ is the orbital semi-major axis of the planet, $q$ is the planet:star mass ratio, $\Omega_p$ is the angular frequency of the planet and $h$ is the disc aspect ratio, and $C$ is an order-unity constant.  The torque scales with the square of the planet mass, so the Type I migration rate $da/dt$ increases linearly with planet mass $M_{\mathrm p}$.  We follow other authors \citep[e.g.,][]{Lee2002,Rein2012}, and parametrize the migration rate as
\begin{equation} \label{da}
\frac{da}{dt} = -\frac{a}{\tau (M_{\mathrm p})} \, .
\end{equation}
In contrast to previous work, we define the migration time-scale to be a function of planetary mass
\begin{equation}
\tau(M_{\mathrm p}) = \tau M_{scale}/M_p  \, .
\end{equation}
Here $\tau$ and $M_{scale}$ are reference values: we choose $M_{scale} = 3 \times 10^{-5} M_{\odot}$ (approximately $10M_{\oplus}$), and treat the migration time-scale $\tau$ as an input parameter. We implement migration (and eccentricity damping; see below) in our code following a method described in the Appendix \ref{integrator}. This method differs from that of \cite{Lee2002} in that the damping is implemented directly as forces at each time-step, and is similar to the method implemented by \citet{Rein2012Rebound}. This approach captures the most important features of Type I migration (the migration rate and scaling with planet mass) while remaining computationally inexpensive. In adopting this method, we assume that all planets migrate inwards only. Several authors have recently shown that outward migration is possible for super-Earth sized planets, and that this behaviour can be important in preventing the loss of planets on to the star \citep{Bitsch2013,Bitsch2014,Cossou2014}. We note however that the only the most massive planets (those with $M_p \gtrsim 5 M_{\oplus}$) considered here are large enough to potentially migrate outwards in the disc models considered by these studies, and even then only in specific regions of the disc \citep[e.g.,][]{Bitsch2014a}. We consider a range of migration time-scales $\tau = 10^{3.5}$--$10^{5.5}$yr, except in the case of Kepler-11. It was clear from early work with this system that a longer migration time-scale was preferred, and hence for Kepler-11 we consider $\tau = 10^{4}$--$10^{6}$yr. We do not base our choice of time-scale on a specific or evolving disc model; instead we explore a broad range of the relevant physical parameters, without reference to the disc properties. It is however possible to relate the migration time-scales presented here to canonical disc models using simple arguments. For instance, for a 1M$_{\odot}$ star and typical parameters ($\Sigma = 1000$g\,cm$^{-2}$, $a=1$AU, $h=0.05$, $M_p = M_{scale}$), equation \ref{eq:type1} gives a migration time-scale $\tau = 6.3 \times 10^4$yr.  Our adopted range of migration time-scales therefore corresponds approximately to a range of disc surface densities $\Sigma \simeq 10^2$--$10^4$g\,cm$^{-2}$ at 1AU.

For low eccentricities ($e \lesssim 2h$) in the Type I regime, both analytic arguments and numerical arguments find that planet-disc interactions lead to exponential damping of the planet's eccentricity \citep{Tanaka2004, Cresswell2007, Bitsch2010}.  We apply eccentricity damping in a similar fashion to that described above, with the damping rate given by
\begin{equation}\label{de}
\frac{de}{dt} = -\frac{Ke}{\tau (M_p)} \, .
\end{equation}
We therefore assume that the eccentricity damping rate $de/dt$ is proportional to the migration rate $da/dt$; $K$ is the constant of proportionality.  This assumption is motivated by several theoretical studies.  \citet{Tanaka2004} used a linear analysis for low-mass planets to show that $\tau_{ecc} \approx (H/r)^2 \tau_{mig}$, suggesting a $K \sim 10^2$ for canonical disc models. Similarly, \cite{Lee2002} found that a value in the range $K = 10$--100 is required to form the GJ 876 system (although the planets in GJ 876 are notably more massive than those considered here). Hydrodynamical simulations performed by \citet{Cresswell2007} and \citet{Bitsch2010} both show excellent agreement the results of \citet{Tanaka2004} in the low $e$ regime, finding eccentricity damping time scales to be a few tens of orbits.  We treat $K$ as an input parameter which sets the strength of the eccentricity damping, and and consider values in the range $K=10^{0.5}$--$10^{2.5}$. Note that no inclination damping is included in these simulations, since all planets are initialised on co-planar orbits.

\subsection{Disc turbulence}
Protoplanetary discs are turbulent, and in addition to transporting angular momentum and driving accretion, this turbulence results in stochastic fluctuations in the local gas density.  These local density variations lead in turn to stochastic variations in the planet-disc torque, which cause the planet's orbit to undergo a random walk \citep{Nelson2004, Nelson2005, Oishi2007}. Full magnetohydrodynamic simulations of planet-disc interactions in a turbulent disc remain extremely computationally expensive, with some authors choosing to apply a stochastic potential to the disc to perform such simulations in an efficient manner \citep[e.g.,][]{Laughlin2004,Baruteau2010,Pierens2012}. Previous work has shown that this process can also be well-approximated in N-body calculations by applying stochastic forcing to the planets. Following the method of \cite{Rein2009}, we use a
modified discrete time Markov process to generate stochastic forces in both the
$\phi$ and $r$ direction at every point along the planet's orbit, hence adding noise to the acceleration of each planet and sending its orbital elements on a random walk (around the smooth net migration rate described in Section 2.1). The standard Markov process is zero-mean and Gaussian, and is defined by two parameters: the root-mean-square (RMS) force-per-unit-mass $\sqrt{\langle F^2 \rangle}$ and the auto-correlation time $\tau_c$.  The RMS force is a free parameter in our model, whilst the auto-correlation time is set to $\Omega^{-1}$ for each planet. We note that this is a high estimate of $\tau_c$ and that numerical simulations show that it can be as low as $0.5\Omega^{-1}$ \citep{Oishi2007}. We characterise the strength of the stochastic forcing in terms of the dimensionless parameter $\beta = \sqrt{\langle F^2 \rangle}/F_g$, where $F_g$ is the gravitational acceleration due to the star's gravity. This means that the absolute magnitude of the stochastic forces grows as planets move inwards, so strictly our distribution of forces is no longer Gaussian (though it remains zero-mean). We vary $\beta$ between $10^{-5}$ and $10^{-8}$ in all simulations presented here.

Our parametrisation of the turbulence can be compared to previous works for scaling. \cite{Paardekooper2013} parametrize their noise according to the scale $F_0 = \pi G \Sigma / 2$, with $\Sigma = 1.5 \times 10^4$g\,cm$^{-2}$, varying  $\sqrt{\langle F^2 \rangle}/F_0$ between 0.01 and 0.1. The corresponding range for our $\beta$ parameter (required to reproduce the absolute magnitude of the forces) is $\beta = 2.5 \times 10^{-7}$ to $2.5\times 10^{-6}$ at 0.1AU.

\subsection{Simulation set-up} \label{setup}
Our simulations have three free parameters: the migration time-scale $\tau$, the eccentricity damping constant $K$, and $\beta$, which characterises the strength of the disc turbulence.  Predicted values of all three of these parameters span at least 2--3 orders of magnitude, so in order to span this vast parameter space we perform 10,000 simulations of each system, each with randomly chosen values of $\tau$, $K$ \& $\beta$.  The values of each parameter are sampled uniformly in log-space in order to understand how order of magnitude changes in their values affect the evolution of each system.

We apply our model to three well-known systems (Kepler-11, -32 and -80), the parameters of which are given below.  We initially place all planets on circular, co-planar orbits with randomly chosen orbital phases. All planets are present at the start of the simulation and migrate together -- the consequences of which we discuss briefly in Section \ref{sec:limitations}. We expect that planets form beyond the snow-line, so we place initially place the inner-most planet in each system at $a=1$AU (a typical location for the snow-line in protoplanetary discs; e.g., \citealt{Garaud2004}).  Beyond this we place the remaining planets in their observed order, with separations motivated by an oligarchic spacing argument \citep[e.g.,][]{Kokubo1998}.  In this scenario protoplanets form with separations of 5--10$r_H$, where $r_H$ is the mutual Hill radius of the two planets
\begin{equation}
r_H = \bigg{(}\frac{M_1 + M_2}{3M_\star}\bigg{)}^{1/3} \frac{a_1 + a_2}{2} \, .
\end{equation}
We note, however, that for the masses considered here this typically places planet pairs inside the 2:1 mean-motion resonance.  Observations suggest that many planet pairs are captured into this resonance, and as we expect most of our planets to undergo convergent migration we instead require a somewhat wider initial spacing.  We assign initial separations by randomly sampling a Gaussian distribution, with mean $30r_H$ and standard deviation $5r_H$ .  This results in most, but not all, planet pairs starting off outside the 2:1 resonance.  This setup is essentially the simplest set of realistic initial conditions that could plausibly reproduce the observed planetary systems.  We recognise, however, that our choice of initial conditions is (necessarily) rather arbitrary, and discuss the consequences of these choices in Section \ref{alt}.  Note also that because the separations, initial phases and turbulent forcing are all randomly chosen, simulations with the same free parameters can have very different outcomes.  Consequently a large number of runs is required to characterise the parameter space adequately.

We halt our simulations when one of four criteria are met: i) a planet is ejected from the system; ii) two planets physically collide; iii) the simulation time exceeds $15\tau$; or iv) a planet attains a semi-major axis that is less than that of the innermost observed planet in a system. The first two criteria represent failed models (i.e., runs which do not reproduce the observed system architectures). Due to the tight packing and high multiplicity of these systems, typically a large number (approx. 70\%) of simulations are ended for this reason.  The third criterion (a time-limit of 15$\tau$) accounts for the fact that a small number of runs evolve into configurations where little or no migration occurs (usually where resonant torques dominate over migration); systems which satisfy this criterion have essentially stopped evolving.  The final stopping criterion is deemed to be a ``successful'' outcome, inasmuch as the innermost planet has migrated to its observed position before any ejections or collisions occur in the system. This stopping criterion is somewhat arbitrary, but there is good reason to believe that the migration of super-Earths might be stopped at small orbital radii in real discs \citep[e.g.,][]{Masset2006}. We discuss the possible physical origins of such a stopping mechanism in further detail in Section \ref{sec:stopping}.

\subsection{Simulations}\label{sec:model}
We use the method presented above to model the assembly of 3 systems: Kepler-11, Kepler-32 and Kepler-80. The parameters we adopt for each of these systems are given in Table \ref{params}.  We summarise the properties of each individual system below.

\subsubsection{Kepler-11}
Kepler-11 is the prototype tightly-packed planetary system, and consists of six super-Earth- to Nepture-size planets orbiting an approximately solar-mass star \cite{Lissauer2011}.  All six (known) planets orbit within 0.5 AU of the host star. The innermost 5 planets are within the orbit of Mercury, having periods between 10 and 47 days, and eccentricities confirmed by dynamical studies of less than 0.02 \citep{Lissauer2013}. These planets also appear to exhibit no mean-motion resonances, although the innermost pair of planets is near to the $5:4$ resonance. The outermost planet is somewhat anomalous, due to its large separation from the others, and its properties are less well constrained.

We adopt the parameters for this system from \cite{Lissauer2013}, who used dynamical fits to the observed transit timing variations (TTVs) to determine the planetary masses. The inner four planets are mass-ordered, with more massive planets orbiting at larger distances from the star, while the fifth planet is less massive.  The mass of the outermost planet (Kepler-11g) is not well constrained by the TTVs, and the dynamical analysis of \cite{Lissauer2013} yields only a weak upper limit to its mass.  In our models this large value leads to unrealistically rapid migration of the outer planet, so we instead adopt a mass of 8$M_{\oplus}$ for planet g, as assumed in the dynamical fitting models in \cite{Lissauer2013}.

\subsubsection{Kepler-32}
Kepler-32 is one of the most compact multi-planet systems discovered to date, with five planets in orbital periods ranging from 0.7--22.8d \citep{Swift2013}.  It also exhibits what appears to be an interlocking mean-motion resonance; planets e and b are apparently in (or very close to) a 1:2 resonance whilst planets b and c are similarly close to a 2:3 resonance. The planets are well ordered in planetary radius, with the planets becoming progressively larger further away from the star.  We adopt the stellar parameters, orbits and planetary radii from table 3 of \cite{Swift2013}. However, in this case the masses of the planets have not been measured directly. We therefore infer masses from the measured planetary radii, using a simple power-law scaling
\begin{equation}\label{scaling}
\frac{M_p}{M_{\oplus}} = \bigg{(}\frac{r_p}{r_{\oplus}}\bigg{)}^{2.06},
\end{equation}
derived from a fit to the Earth and Saturn \citep{Lissauer2011a}. There are likely to be large inaccuracies in these mass estimates: \cite{Lissauer2013} note that even after reducing their size estimates, the planets in the Kepler-11 system are all significantly less massive that one would estimate from this relation.

\subsubsection{Kepler-80}
Kepler-80 (also known as KOI-500) is an extreme example of a tightly-packed system, with four planets in orbital periods that range from just 3.0--9.5d. As with Kepler-32 the planets are well ordered in size, and there is a possible 4-body interlocking resonance of $4:6:9:12$ between the outer four planets \citep{Xie2013}. Parameters for this system are drawn from the NASA Exoplanet Archive (http://exoplanetarchive.ipac.caltech.edu), apart from planet f, whose period and radius are taken from \cite{Xie2013}. Only radii and orbital periods are available for the planets of Kepler-80, so we derive semi-major axes from the orbital periods (using the provided stellar mass), and again use the scaling relation (equation \ref{scaling}) to obtain mass estimates for the planets.

\begin{table}
\begin{center}
\begin{tabular}{r c c c}
\hline \hline
Object & M ($M_{\odot}/M_{\oplus}$)& a (AU) & r ($R_{\odot}/R_{\oplus}$) \\ \hline \hline
\textbf{Kepler-11 *} & 0.961 & --- & 1.053 \\ \hline
Kepler-11 b & 1.9 & 0.091 & 1.8 \\ \hline
Kepler-11 c & 2.9 & 0.107 & 2.87 \\ \hline
Kepler-11 d & 7.3 & 0.155 & 3.11\\ \hline
Kepler-11 e & 8.0 & 0.195 & 4.18 \\ \hline
Kepler-11 f & 2.0  & 0.25  & 2.48 \\ \hline
Kepler-11 g & 8.09 & 0.466 & 3.33 \\ \hline \hline
\textbf{Kepler-32 *} & 0.57 &  ---   & 0.53 \\ \hline
Kepler-32 f & 0.65 & 0.013  & 0.81 \\ \hline
Kepler-32 e & 2.31 & 0.0323 & 1.50 \\ \hline
Kepler-32 b & 5.07 & 0.0519 & 2.20 \\ \hline
Kepler-32 c & 4.17 & 0.067  & 2.00 \\ \hline
Kepler-32 d & 7.74 & 0.128  & 2.70  \\ \hline \hline
\textbf{Kepler-80 *} & 0.72  &  ---  & 0.637 \\ \hline
Kepler-80 f & 1.46 & 0.017 & 1.2 \\ \hline
Kepler-80 d & 2.31 & 0.037 & 1.5 \\ \hline
Kepler-80 e & 2.63 & 0.049 & 1.6 \\ \hline
Kepler-80 b & 7.16 & 0.065 & 2.6 \\ \hline
Kepler-80 c & 8.34 & 0.079 & 2.8 \\ \hline \hline

\end{tabular}
\caption{Simulation parameters for each of our 3 systems. Units relative to the sun are used for stellar components of systems, units relative to earth are used for their planetary companions. See section \ref{setup} for a detailed description of the origin of these values.} \label{params}
\end{center}
\end{table}
\section{Results}
\begin{table}
\begin{center}
\begin{tabular}{ r c c c c c c }
\hline
System& S & S/O & S/U & T & C & E \\ \hline \hline 
Kepler-11 & 3270 & 2089 & 1181 & 1& 5753 & 977\\ \hline
Kepler-32 & 1964 &1916 &  48 & 153& 7140 & 896\\ \hline
Kepler-80 & 2106 & 2031 & 75 & 40 & 6907 & 987 \\ \hline

\end{tabular}
\caption{The number of runs ending in a particular outcome for each set of 10,000 runs. S: Runs that finished without a collision or ejection event. S/O: Subset of S that finished with the planets correctly ordered. S/U: As S/O but for incorrectly ordered planets. T: Total number of runs from S that were stopped due to running for longer than $15\tau$. C: Total number of runs ending in a collision of 2 bodies. E: Total number of runs ending in the ejection of a body. \label{outcomes}}
\end{center}
\end{table}

\begin{figure*}
\begin{center}
\includegraphics[width=\linewidth]{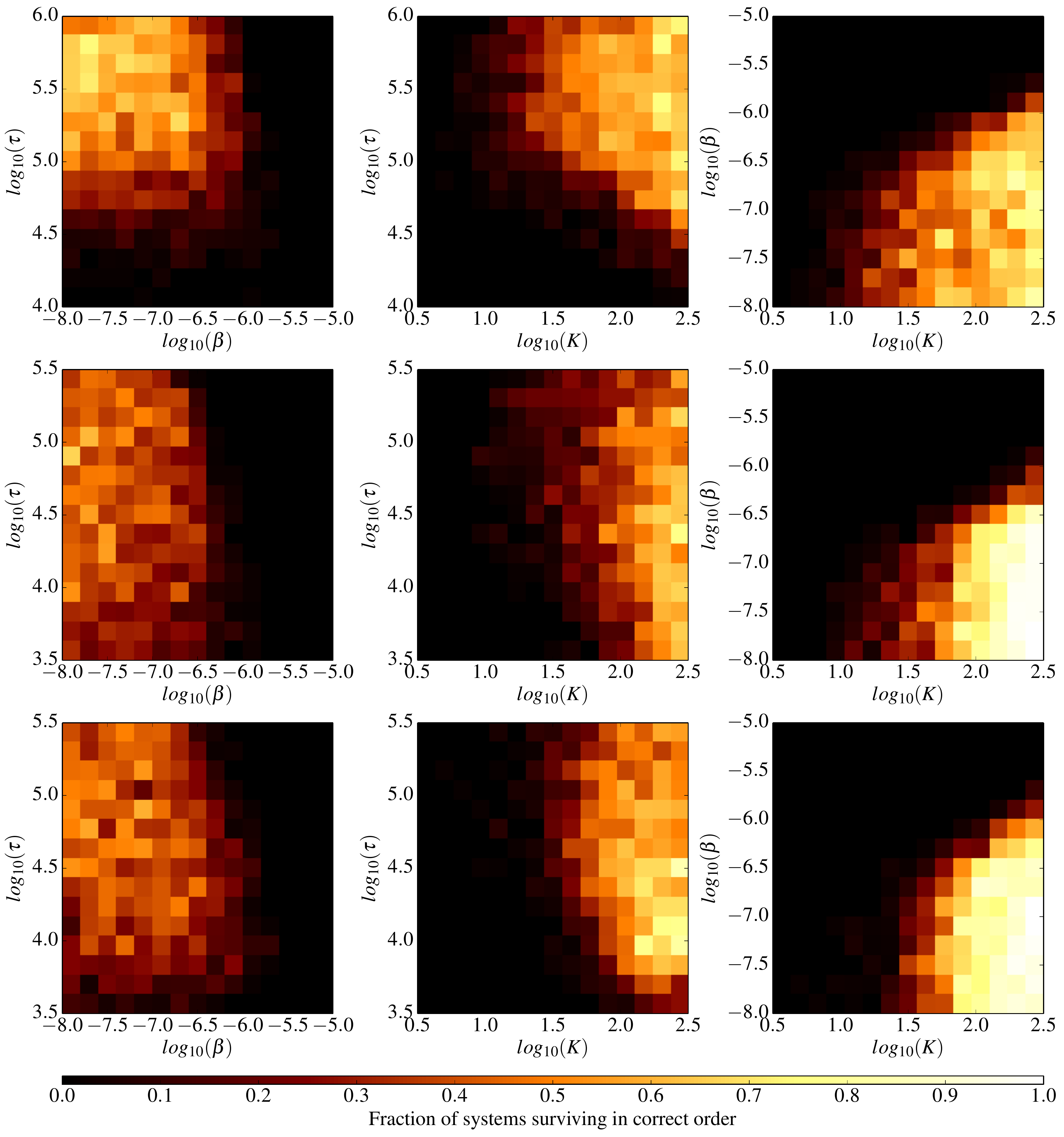}
\caption{Fractions of successful runs as a function of our free parameters $\tau$ (migration time-scale), $K$ (eccentricity damping) and $\beta$ (``turbulent'' forcing).  Top row: Kepler-11; middle row: Kepler-32; bottom row: Kepler-80.  Our calculations generally favour strong eccentricity damping and weak to modest levels of stochastic forcing, but do not show a strong dependence on migration rate.
 \label{kep11params} }
  \end{center}
\end{figure*}
\begin{figure*}
\begin{center}
\includegraphics[width=\linewidth]{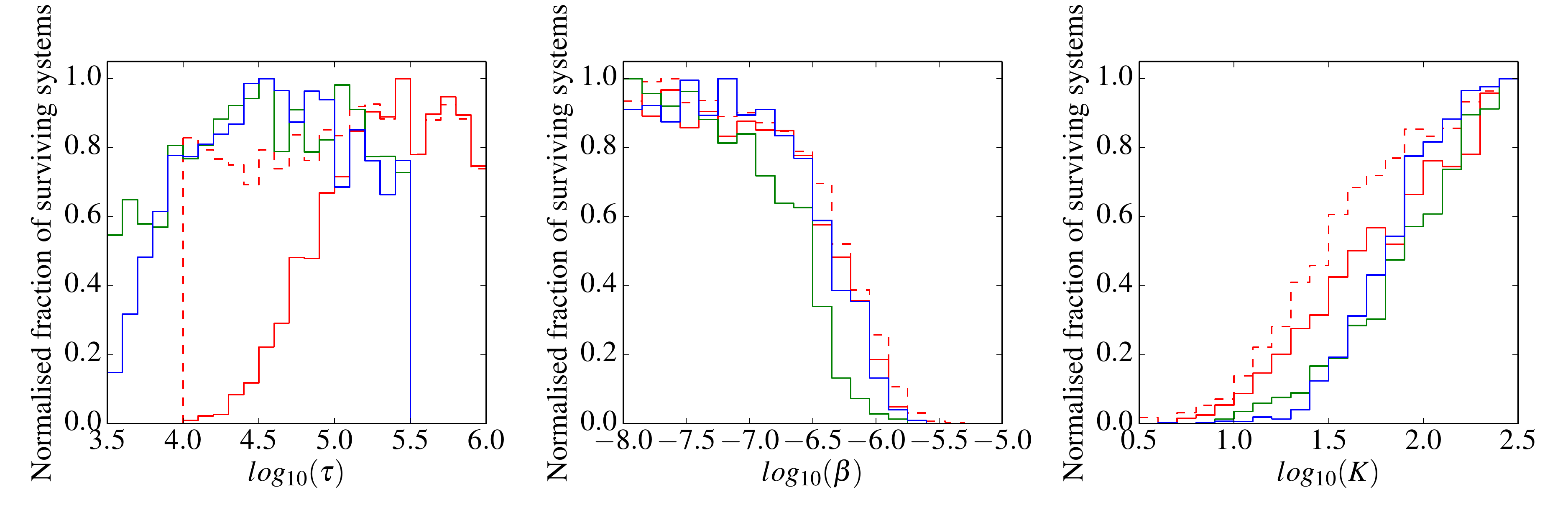}
\caption{Historgrams showing success rate for each region of flattened parameter space across all simulations. Red: Kepler-11, dashed-red: as with red line but alllowing planets 5 and 6 to maintain order or switch positions with one another, green: Kepler-32, blue: Kepler-80. The vertical axis shows the fraction of systems run in each bin that fulfilled the success criteria, normalised to a maximum of 1 for easy comparison. The shapes of the distributions are very similar for all three sets of models. \label{fig:all_1d} }
  \end{center}
\end{figure*}
\begin{figure}
\begin{center}
\includegraphics[width=\linewidth]{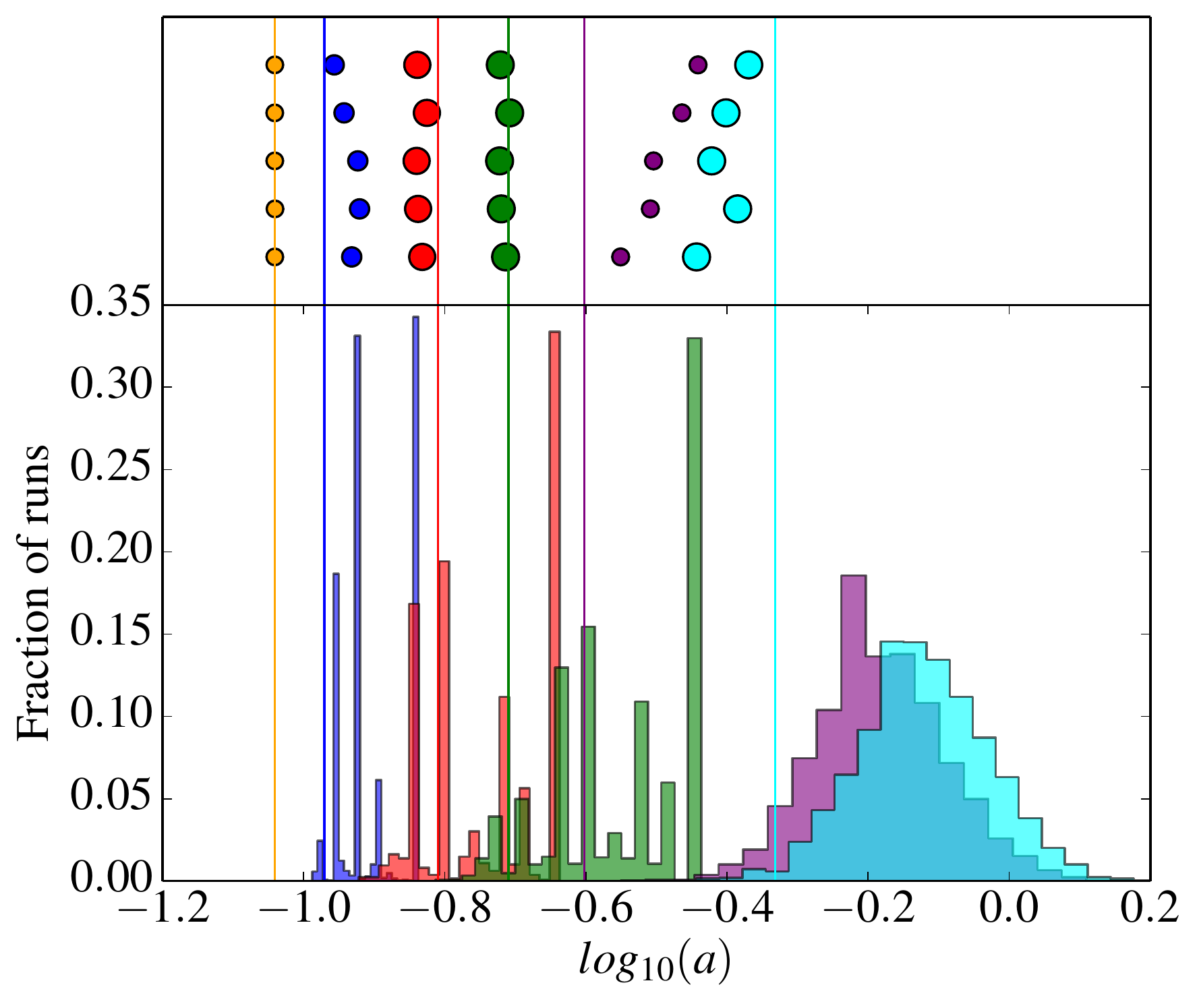}
\includegraphics[width=\linewidth]{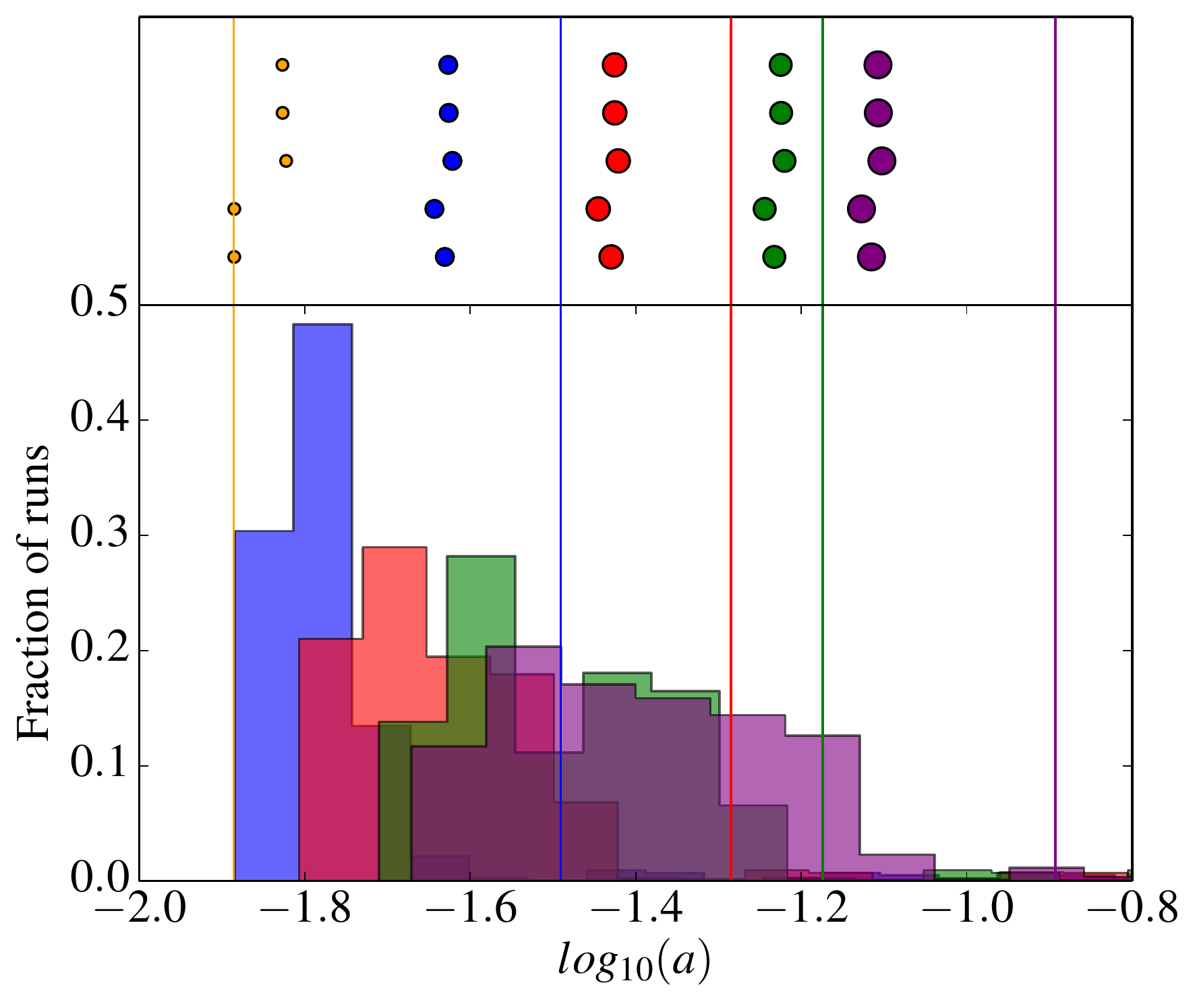}
\includegraphics[width=\linewidth]{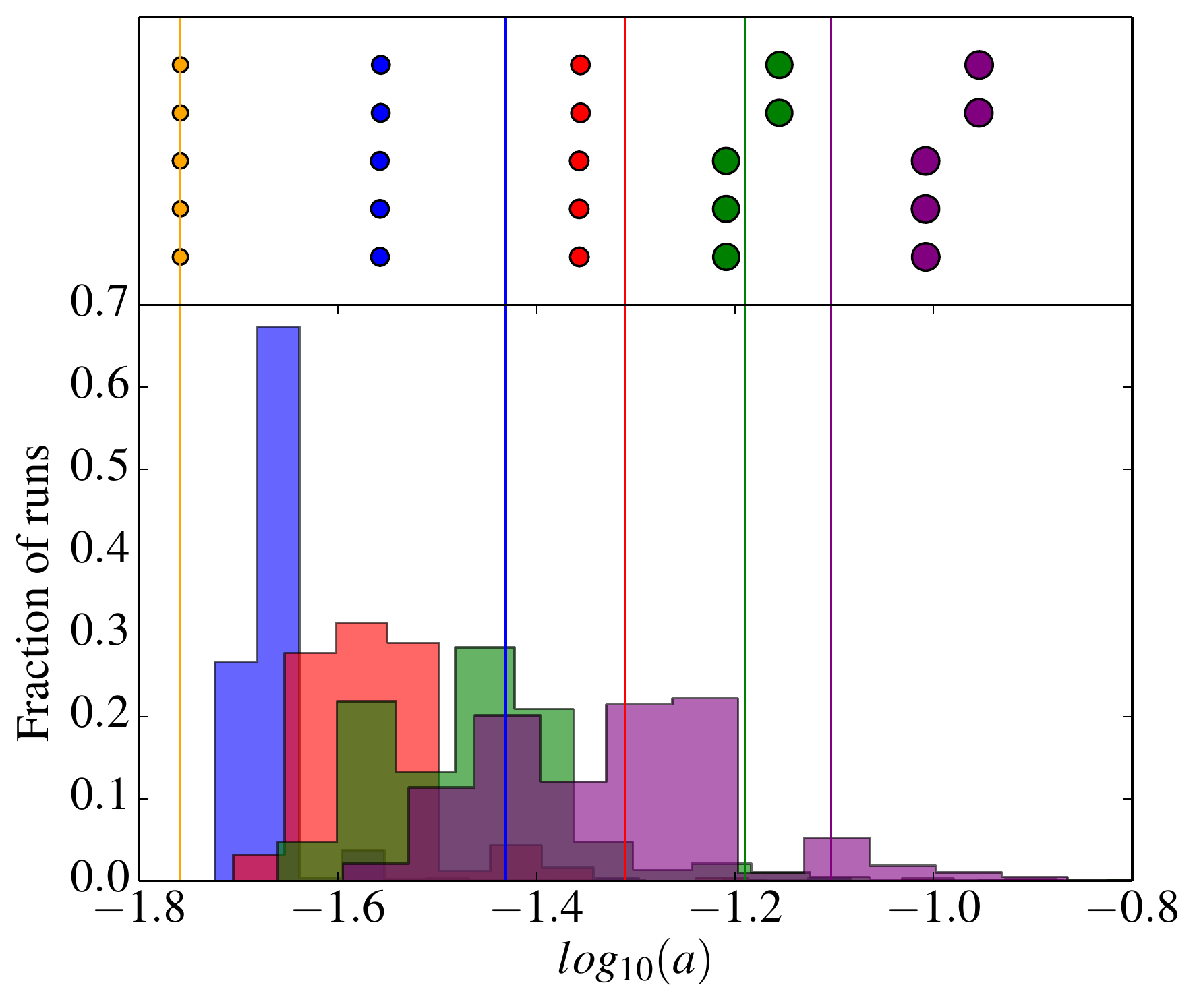}
\caption{Histograms showing distribution of each planet in each system across all successful runs.  Each colour represents a specific planet in a specific system. Solid lines show the actual positions of the planets in each system, whilst points plotted above show planetary positions in some representative "best-fit" models, fitted by semi-major axis. From top to bottom: Kepler-11, Kepler-32, Kepler-80. Note that a small fraction of models are cut off the right of each plot. \label{kep11finala} }
  \end{center}
\end{figure}
\begin{figure}
\begin{center}
\includegraphics[width=\linewidth]{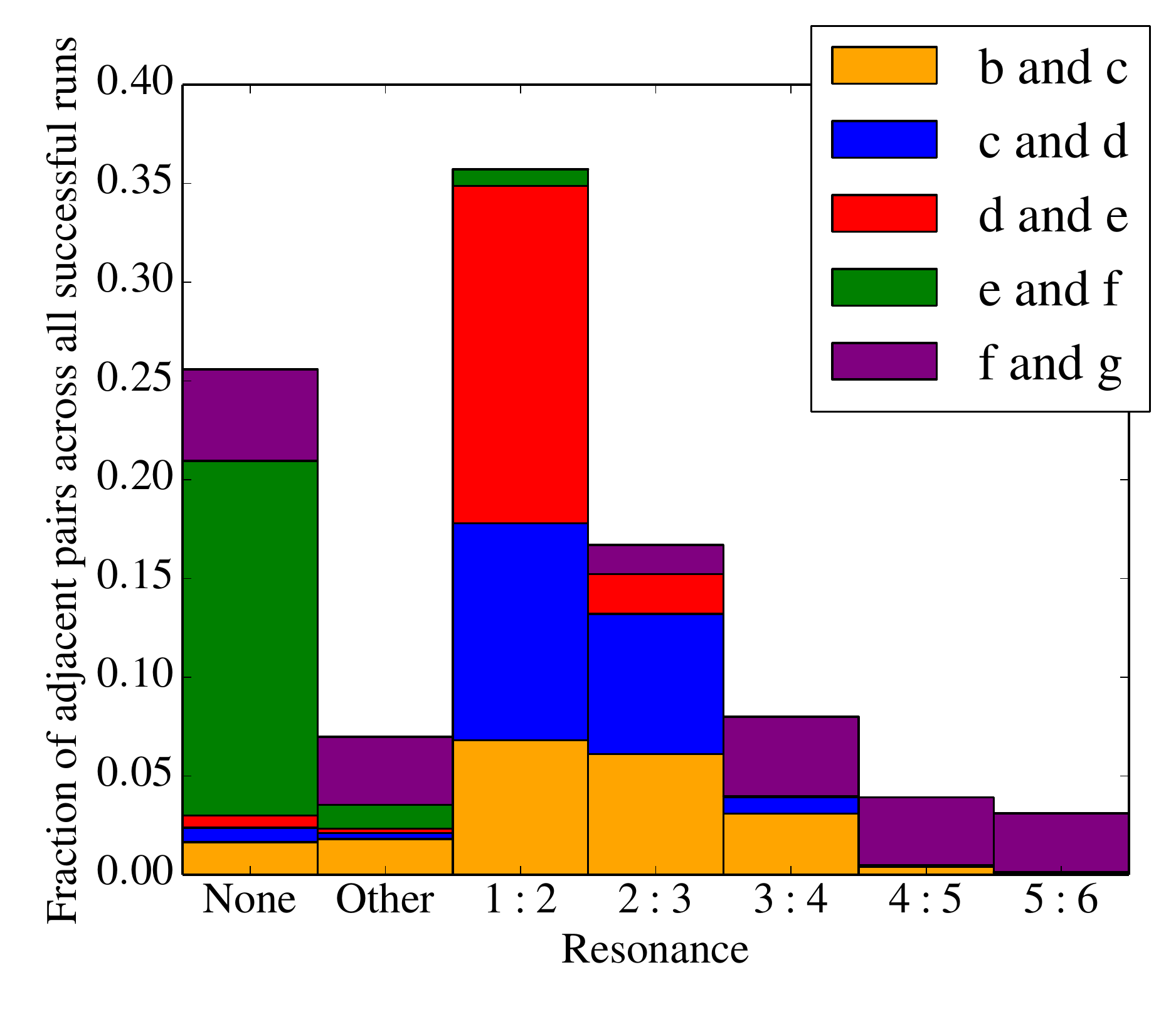}
\includegraphics[width=\linewidth]{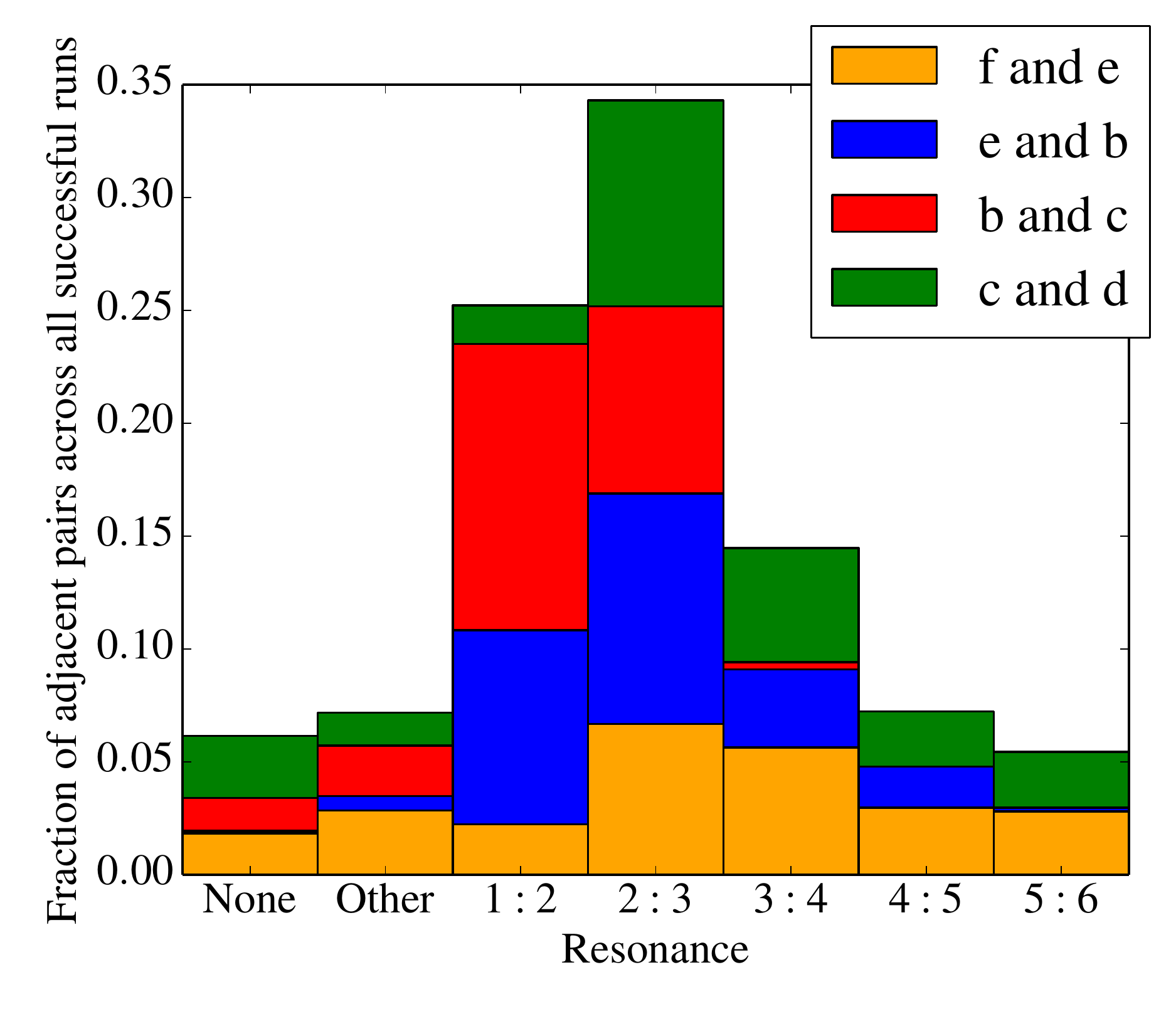}
\includegraphics[width=\linewidth]{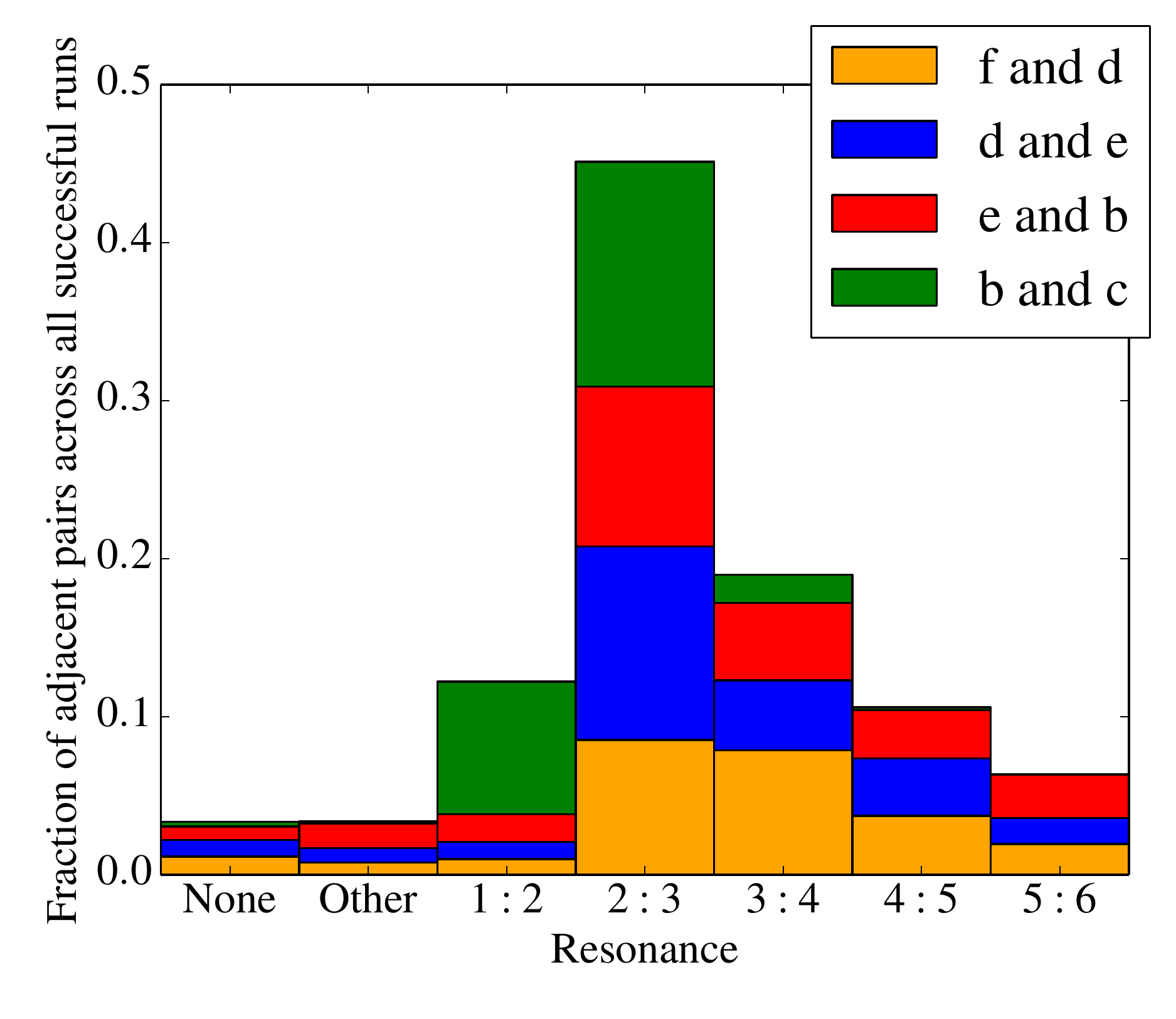}

\caption{Fractions of adjacent pairs of planets that are in each resonance at the end of our successful runs for Kepler-11 (top), Kepler-32 (middle) and Kepler-80 (bottom). Pairs are numbered in order of increasing semi-major axis. The "Other" bar is a sum over all resonances that contain less than 3\% of the total number of pairs. The letters in each key refer to which 2 adjacent planets are represented by each colour.\label{resonancehist} }
  \end{center}
\end{figure}

As described in Section \ref{sec:model}, we ran 10,000 models for each of the three systems.  Given the chaotic nature of the models, and our arbitrary choice of initial conditions, we initially take a very simple view of the results. We define a model as ``successful'' if the innermost planet migrates to its observed position or the simulation time reaches $15\tau$ without any ejections or collisions occurring in the system.  We further split this group of successful runs into two, depending on whether or not the order of the planets was preserved or not.  The numbers of successful/unsuccessful runs for each system are given in Table \ref{params}, while Figures \ref{kep11params} \& \ref{fig:all_1d} show how the fraction of successful runs for each system varies as a function of the input parameters.

While the overall fraction of successful models is rather low, Figure \ref{kep11params} shows that this is essentially an artefact of the large parameter space we consider; in each case there is always a region of parameter space where most of the models are successful.  It is immediately clear from Figures \ref{kep11params} and \ref{fig:all_1d} that our models show a clear preference for stronger eccentricity damping and lower stochastic forcing. The reason for this is clear: the close packing of the planets means that even small perturbations can lead to orbit crossing and subsequent collisions or ejections. A lower level of stochastic forcing reduces the probability of orbit crossing, as does more efficient eccentricity damping. All three sets of models require $\beta \lesssim 10^{-6.5}$, suggesting that modest or low levels of turbulence are required in discs that form compact systems. Similarly, we require the eccentricity damping parameter $K \gtrsim 10^{1.5}$ in order to form tightly packed systems effectively. This is towards the upper end of the range of values found in previous numerical calculations, as discussed above, but is certainly plausible for planets in the Type I migration regime.

When we consider the migration time-scale $\tau$, however, there is a distinct difference in the results for Kepler-11 compared to the other two systems. Our models of Kepler-32 and -80 models show no strong variations with $\tau$: the fraction of successful runs is essentially constant across the range of migration time-scales we consider, and shows only a weak decline for very short migration time-scales ($\tau < 10^4$yr). By contrast, for Kepler-11 we find a strong preference for $\tau \gtrsim 10^5$yr, and see essentially no successful models with $\tau \lesssim 10^{4.5}$yr. 
 
On closer inspection, however, we find that the preference for longer migration time-scales in Kepler-11 is almost entirely due to the ordering of the outer two planets (f \& g). Kepler-11 is the only system we consider where the planets are not ordered in increasing size. If we relax our criteria to include models where planets f \& g are allowed to switch positions (see the dotted histogram in Figure \ref{fig:all_1d}), then we again find that the fraction of surviving systems is approximately constant with $\tau$, as for Kepler-32 and -80.  This behaviour is again readily understood.  In our models planet g is approximately four times more massive than planet f, so the planets' orbits converge rapidly as they migrate. If migration is relatively slow then the probability of capture into a mean-motion resonance is high \citep[see e.g.,][]{Mustill2011}, but resonant capture becomes progressively less likely for faster migration rates, making it much more likely that planet g ``overtakes'' planet f.  The mass of planet g is poorly constrained by current observations \citep{Lissauer2013}, however, so it is not clear whether this result is significant.  Otherwise, we find that migration time-scales $\tau \simeq 10^4$--$10^6$yr readily lead to successful assembly of these compact systems.  This range of migration time-scales is broadly consistent with the predictions of Type I migration models, but in general our calculations do not set strong constraints on the required migration rate.

Figure \ref{kep11finala} shows the distribution of final semi-major axes for each planet in each system at the end of our successful runs. Also shown are some ``best-fit'' models, in which the final positions of the planets are close to their observed locations. Generally the systems produced by our models are indeed very tightly packed, with the majority of systems in the Kepler-11 case containing all 6 planets within 1AU. In most cases our models end with the outer 3 planets in the Kepler-11 system exterior to their true positions, but even in these cases the tightly-packed nature of the system is clearly maintained.  By contrast, our models of Kepler-32 and Kepler-80 tend to produce systems that are in fact more tightly-packed than the real systems, particularly with regards to the spacing of the first and second planets.  Thus, although the precise orbital configurations of the observed systems are somewhat unusual outcomes, our models show that it is clearly possible to assemble systems of five or more tightly-packed at sub-AU radii through simultaneous, disc-driven migration 

\subsection{Mean-motion resonances} \label{resonances}
Further insight into our results can be obtained by considering the occurrence of mean-motion resonances in the successful runs. Calculating the resonant argument for multiple planet pairs in thousands of individual simulations is computationally expensive, so we instead perform a simple analysis on the period ratios in the final orbital configuration of each run to establish if planets are in resonance. We take the ratio of periods between each pair of adjacent planets, find the closest integer ratio, and consider it a possible resonance actual ratio is within 0.5\% of this integer value. We acknowledge that resonant behaviour can be observed even between planets that are $\sim 5\%$ away from the exact commensurability \cite[see e.g.,][]{Raymond2008a}. However, we found that for our low-eccentricity systems the $0.5\%$ tolerance was sufficient, since resonant pairs tend to remain very close to exact commensurability, while non-resonant pairs are typically very far from the nearest first-order resonance. This analysis is performed in each of the final 10 output snapshots from each run (i.e., over {100 yr}), and the planets are considered to be in resonance if the nearest integer ratio is the same across all 10 snapshots, and always falls within the 0.5\% tolerance.

We find that vast majority of runs end with each pair of adjacent planets in resonance, as shown in figure \ref{resonancehist}. However, non-resonant configurations are by not uncommon: for each pair of adjacent planets in Kepler-11, a few percent of runs end with no resonance. In our Kepler-11 models the outer two planets are almost invariably trapped in one of several different resonances at the end of the simulations. This is a result of the strongly convergent migration described above, and partly explains the difficulty our models have in keeping these planets as well spaced as they are in reality. Moreover, this same mechanism explains the difficulty in spacing planets e and f correctly: planet f is completely dominated by its resonant interaction with g, allowing planet e to migrate inwards unperturbed. This in turn accounts for the dearth of resonances between planets e and f seen in figure \ref{resonancehist} - these being the only two planets that are outside of a resonance in the majority of our runs. However, given the large uncertainty in the mass of planet g (discussed above), the significance of this behaviour is unclear.

In general, our models of Kepler-32 and -80 show similar behaviour to Kepler-11, with the caveat that in reality a number of (possible) mean-motion resonances are observed in these systems.  In both of these systems the period ratio of the innermost pair of planets at the end of our simulations is significantly smaller than in reality (see Fig.\ref{kep11finala}), but otherwise our models have some success in reproducing the observed MMRs in these systems. In Kepler-32 34\% of successful runs end with planets e and b in a $2:1$ resonance, and 33\% with planets b and c in a $3:2$ resonance; the 1:2:3 resonance in Kepler-32 occurs 52 times in 1964 successful runs (i.e., 2.6\% of the time).  In Kepler-80, the interlocking resonance between planets d through c occurs only once in our successful runs, though the individual resonances occur separately many times. The 3:2 resonances between planets d and e and planets e and b occur 49\% and 40\% of the time respectively, whilst the 4:3 resonance between planets b and c occurs 20\% of the time.  

The prevalence and population of resonances depends weakly on the model parameters. For Kepler-11 the innermost pair of planets is twice as likely to be found out of resonance for values of $\beta$ above the median (10.8\% of successful runs) than below (5.7\%). \cite{Rein2012} similarly found that increased stochastic forcing could reduce the number of observed resonances in systems. This effect is not as pronounced in our results as in those of \cite{Rein2012} however, with the majority of our planet pairs still found in first order resonances even in simulations with higher-than-median values of $\beta$. When we consider the dependence on the migration time-scale $\tau$, we see similar behaviour to that found by \cite{Paardekooper2013}: faster migration leads to planets crossing first-order resonances such as $2:1$ and $3:2$, and instead becoming trapped in small-period ratio first-order resonances such as $4:3$ and $5:4$. For instance, in Kepler-32, the innermost pair of planets are found in the $3:2$ resonance at the end of 22\% of successful simulations for values of $\tau$ above the median. For values below the median, just 4\% appear to be in this resonance, with the corresponding percentage for the $4:3$ resonance jumping from 7\% to 15\%. A similar trend is seen in the other two systems. Additionally, we find that $\tau$ can affect the percentage of planet pairs that finish the simulation in no resonance. In Kepler-32 simulations with $\tau$ larger than the median, 1\% of all planetary pairs are in no perceivable resonance, this figure jumping to 5.2\% for values of $\tau$ below the median. However, it is still clear that our models in general over-predict the occurrence of MMRs, and we discuss this issue further in Section \ref{resonances-discussion}.

Interestingly, whilst the overall prevalence of resonances is similar across all three systems, figure \ref{resonancehist} shows that there are marked variations in which resonances are preferred. For instance, over 25\% of adjacent pairs end trapped in the 2:1 resonance in Kepler-32, with less than half this number ending in the same resonance in Kepler-80. This is particularly curious given the striking similarities between these two systems. Our models of both Kepler-32 and Kepler-80 generally favour the $3:2$ resonance, while Kepler-11 favours instead the 2:1 resonance, with the 3:2 resonance occurring much less often. It is clear that the minor differences between these systems (e.g., in planet mass or initial spacing) can make a large difference to the preferred configurations of resonances in our models.

\subsection{Stability}
A key additional consideration is the orbital stability of the configurations -- it is entirely possible that many of our systems will become unstable once the damping and forcing due to the disc is no longer present. Simulations of super-Earth embryos conducted by \cite{Cossou2014} suggest that this indeed may be the case, with the removal of disc-driven damping leading to instability and collisions in their calculations. We performed basic stability tests on the successful Kepler-11 models by taking the final results of each and evolving them using the 4th order integrator described by \cite{Yoshida1990} for 5Myr \footnote{We restrict this analysis to Kepler-11 since we expect the stability of Kepler-32 and Kepler-80 to be dominated by the uncertainty in their planetary masses.}. We consider models to be unstable if they undergo an ejection or collision during this time, or if they show signs of Lagrange instability.  Following \cite{Deck2013}, we define Lagrange instability to be a change in  the semi-major axis of any planet by more than 5\% from its initial value over the course of the 5Myr. Of the 2089 successful Kepler-11 models, 5.03\% underwent an ejection or collision event within the 5Myr integration. A further 0.24\% were found to be Lagrange unstable. Generally we find that the unstable models are those with little eccentricity damping, since they have higher eccentricities at the beginning of the stability test which can quickly lead to orbit crossing. A small percentage (1.82\%) of the models required time-steps which were prohibitively short, and hence were not completed. The remaining 92.92\% of the models were stable for the full 5Myr. This rudimentary analysis shows that the majority of our models are stable on $\gtrsim$ Myr time-scales, and consequently that the stability of our models does not affect our results significantly.

\section{Discussion}
\subsection{Simulation outcomes \& preferred parameter values} 
Table \ref{outcomes} shows the total number of runs that ended in each type of outcome for each system. The tiny fraction of runs which are ended due to the simulation time limit implies that this criterion has little effect on our results, and a manual inspection of the final states of these models reveals that they are generally very close to reaching the normal, position-based stopping criterion anyway. In general, runs which end without a collision or ejection event end with the planets in the same order. Kepler-11 is the notable exception to this trend, with $20.89 \%$ of runs ending successfully and a further $11.50 \%$ ending without any catastrophic event but with planets f and g having swapped positions. The apparent ease with which this planets switch positions suggests that these systems may not have formed in the orders that they are currently observed, particularly given that planet f in Kepler-11 does not follow the otherwise prevalent trend of mass ordering. Instead the planets could have formed in a different order, with the combination of strong damping, convergent migration and disc turbulence facilitating their rearrangement without violent interactions. 

As previously stated, our results set no specific constraints on the migration time-scale required for the assembly of these systems. This suggests that limits on disc parameters are instead set by our allowed values of $K$ and $\beta$, and also means that any realistic Type I migration time-scale allows the assembly of these systems via disc-driven migration. This is convenient since it allows a number of potential migration stopping mechanisms to be invoked in ending the migration of the planets (see Section \ref{sec:stopping}). The eccentricity damping parameter $K$ is much more strongly constrained, with values $K < 10^{1.5}$ almost always ending in a collision or ejection. Stronger damping naturally allows for easier assembly, since the combination of tight-packing and resonant configurations in our results can lead to orbit crossing if left unchecked. In spite of this, the preferred range of $K$ values is broadly consistent with those found in previous studies (see Section \ref{sec:migration}) and thus compatible with typical Type I migration models.

The preferred range of $\beta$ values in our simulations is $\beta \lesssim 10^{-6.5}$, with larger values usually leading to collisions, ejections or planet reordering. To obtain some insight into the meaning of this value, we calibrate our $\beta$ parameter against previous magnetohydrodynamical studies of migration in turbulent discs. For a disc with \cite{Shakura1973} viscosity parameter $\alpha = 0.007$ and aspect ratio $h= 0.07$, \cite{Nelson2004} found the RMS of the fluctuating specific torque to be $\sigma \simeq 2 \times 10^{-5}$. In our units this corresponds to $\beta \simeq 6 \times 10^{-5}$. However, the disc aspect ratio adopted by \cite{Nelson2004} is significantly larger than expected at the sub-AU radii we consider here: for a 1M$_{\odot}$ star, a gas temperature of 1000K yields $h \simeq 0.02$ at $r=0.1$AU. Since the strength of turbulent fluctuations scales approximately as $h^2$, a realistic value of $\beta$ at these orbital radii is likely to be at least an order of magnitude lower: we estimate that the torque fluctuations in the calculations of \cite{Nelson2004} correspond to $\beta \approx 10^{-6}$ at $r \approx 0.1$AU.  This is around the upper limit of suitable values in our simulations, and suggests that modest levels of disc turbulence ($\alpha \approx 10^{-3}$--$10^{-2}$) are compatible with the assembly of compact planetary systems by simultaneous migration.  We note, however, that in real discs the \cite{Shakura1973} $\alpha$-parameter depends on fluctuations in the velocity and magnetic fields \citep[e.g.,][]{Balbus2011}, while the migration torque varies due to fluctuations in the gas density.  Establishing a relationship between the turbulent stresses which drive angular momentum transport and the density fluctuations that give rise to stochastic migration (i.e., between $\alpha$ and $\beta$) therefore requires detailed magnetohydrodynamic calculations, and is beyond the scope of this work.

\subsection{Mean-motion resonances}
\label{resonances-discussion}
One of the principle arguments against forming this class of systems (in particular Kepler-11) via convergent migration is that many of the adjacent pairs of planets within them do not appear to be in MMRs. As discussed above, our models have significant difficulty replicating this dearth of resonances. It is clear from figure \ref{kep11params} that stronger turbulence does not reduce the incidence of resonances significantly, but instead simply reduces the survival rate of systems. An alternative explanation for the paucity of resonances, especially in Kepler-11, is hence required. Mechanisms that could lead to the breakdown of resonances have already been studied in great detail, largely in an attempt to explain an observed pile-up of planetary pairs that have period ratios slightly larger than the closest MMR \citep{Lissauer2011a,Fabrycky2012}.

\cite{Goldreich2014} have suggested that the paucity of resonances observed in exoplanetary systems may be a side-effect of disc-driven eccentricity damping. However, the presence of this effect in our simulations appears to have had little effect on the long-term maintenance of resonances. Given that Kepler-11 is estimated to be 8.1Gyr old \citep{Lissauer2013} and protoplanetary disc lifetimes are of order several Myr, long term dynamical evolution of these systems could play an important role in finalising their architectures. For instance, the excitation of eccentricity in the post-disc phase can cause orbit crossing and close-encounters, leading to the breakdown of resonances \citep{Ida2010}. A cursory analysis of MMRs present at the end of our Kepler-11 stability analysis suggests that this effect is not significant after 5Myr for our models. 

Several authors have investigated the possibility that the paucity of resonances can be attributed to tidal interactions. \cite{Lee2013} find that tidal dissipation can drive some planets out of resonances, but that the effect is generally not strong enough to account for all resonance breaking. \cite{Lithwick2012} show that tidal dissipation alone can only explain the complete distribution of period ratios if the damping is unexpectedly strong. They propose instead that planets can be repulsed from a resonance by a dissipative process such as tidal damping acting on the forced eccentricity which is driven by the resonance. \cite{Baruteau2013} show that planets with periods larger than 10 days are unlikely to be repulsed from resonance by tidal interactions. Instead they propose that the wake generated by a companion planet while the two are migrating in resonance can reverse convergent migration. Another promising possibility was suggested by \cite{Chatterjee2014}, who found that interactions with a disc of planetesimals can lead to migration of pairs of resonant planets. This can in turn disrupt the MMR and leads to the final period ratio between the two planets being slightly larger than that of the initial MMR. Whatever the mechanism, there is clearly a fine balance to be struck between breaking resonances in systems such as Kepler-11, and maintaining them in systems such as Kepler-80.

\subsection{Limitations of the model} \label{alt}\label{sec:limitations}
First, we note here that caution should be used when considering the outcome of models with low $\tau$ and high $K$ values (i.e., those with very rapid eccentricity damping). The simulations of \cite{Cresswell2007} suggest that eccentricity damping occurs on several tens of orbital time-scales, while our formalism assumes only that all effects happen on time-scales greater than the orbital period (see Appendix \ref{integrator}). For models in which the outermost planets are initially placed far from their host star, extreme values of low $\tau$ and high $K$ can result in an initial eccentricity damping time-scale of less than one orbital period. However, this occurs only in a minority of our models, and even our strongest eccentricity damping acts on times-scales no smaller than $\simeq 10\mathrm{yr}$ -- far longer than the orbital time-scale at the final orbital radii. Hence we still consider these models to be valid on the basis that the most important period of migration (when the planets are most tightly spaced) is unaffected.

Naturally, for the sake of this simplified calculation, we have made several assumptions which are not valid for actual planetary systems. For instance, the parametrization used here is evidently not self-consistent; previous studies such as those discussed in Section \ref{sec:migration} have shown that eccentricity damping and migration time-scales are not independent. Given the uncertainties in these values however, it is not possible to perform a more realistic calculation at this time without reverting to full hydrodynamical models. In spite of this, it is encouraging that the range of $\tau$ and $K$ values for which our models are successful is broadly consistent with the results of these more sophisticated calculations.

The initial conditions in our models have been chosen on the basis of simple physical arguments, but the planets considered here most likely did not form concurrently with their current masses. According to the standard core-accretion model of planet formation, these planets would have spent considerable time as lower mass planetary embryos, accreting material from a protoplanetary disc \citep{Raymond2013} and even merging with other embryos. In particular, several planets in Kepler-11 have bulk densities which are consistent with a gaseous atmosphere component, while the planets themselves are sub-Jupiter mass. This suggests that they underwent a phase of relatively slow accretion from the gas disk \citep{Lissauer2013}. If the planets formed at different times, then one would expect that the innermost planets would have formed before their further-out counterparts and hence begun migrating first. This sort of sequential migration could vastly alter the final spacing of the planets, and therefore the landscape of resonances seen in Figure \ref{resonancehist}. Allowing for different formation times and periods of slower migration with lower planetary masses may alleviate some of the difficulties with spacing the planets in our models. Furthermore, we have seen in our results for Kepler-11 that the order of planets in such systems is not necessarily fixed, and allowing the planets to begin in different orders may alter the preferred region of parameter space or the resulting distributions of period ratios and semi-major axes.

\subsubsection{Stopping} \label{sec:stopping}
The largest simplification in the calculations presented here is that planet migration is halted arbitrarily. In reality is is not clear how Type I migration comes to an end, or indeed why it should end so much closer to the host star for these planets than in other planetary systems. There are, however, two mechanisms that are commonly suggested to halt Type I migration, and the large range of possible migration time-scales allowed in our models suggests that either of them may be viable.

The first of these mechanisms is disc clearing \citep[see, e.g.,][and references therein]{Alexander2014}. Protoplanetary disc dispersal is driven by processes (disc accretion and mass-loss due to winds) which are largely independent of planet formation, and removal of the disc gas inevitably halts disc-driven planet migration. The role of disc clearing in halting Type II migration has been studied in detail by various different authors \citep[e.g.,][]{Armitage2002,Alexander2012}, but the implications for the Type I regime are largely unexplored. Typical protoplanetary disc lifetimes are 1--10Myr, but final disc clearing occurs rather more rapidly (in $\sim 10^5$yr). Once disc clearing begins, material at sub-AU radii is simply accreted on to the star, resulting in an exponential decline in the disc surface density (on time-scales $\approx10^5$yr) which simply ``strands'' migrating planets at their current positions. This is therefore a plausible mechanism for halting relatively slow Type I migration (i.e., $\tau \gtrsim 10^4$--$10^5$yr), but is difficult to reconcile with shorter migration time-scales. In addition, disc dispersal at sub-AU radii is essentially scale-free, so (to first order) we do not expect disc clearing to alter the architectures of compact planetary systems dramatically.

The alternative to disc clearing is the presence of ``traps'' in the disc: sharp radial changes in the disc structure which result in locations at which an embedded planet experiences no net torque. Such traps may occur at the edges of of a dead zone in an partially ionized disc \citep[e.g.,][]{Gammie1996}, where the inner disc is truncated by the stellar magnetic field \citep[typically at a few stellar radii; e.g.,][]{Hartmann1994,Bouvier2007} or at the dust sublimation radius \citep[which occurs at $r \approx 0.2\mathrm{AU}$; e.g.,][]{Eisner2005}. This mechanism is more plausible for the shorter migration time-scales that we consider ($\tau \lesssim 10^4 \mathrm{yr}$), when the migration phase is less likely to overlap with the end of the disc lifetime. Moreover, as traps occur at specific locations in the disc, they are likely to alter the architectures of migrating planetary systems significantly. \cite{Masset2006} show via hydrodynamical simulations that this mechanism is effective for planets in the super-Earth to sub-Neptune regime out to 5AU in an MMSN-style disc. The aforementioned migration simulations by \cite{Terquem2007} invoked magnetospheric truncation of the disc as a method of halting migration. \cite{Ida2010} conduct population synthesis models of super-Earth systems and find that a magnetospheric cavity in the disc can halt the migration of planetary embryos as they approach the disc edge at around 0.1AU, leading to multiple super-Earths in short-period orbits. Planet-planet interactions may then break any resonances that have formed. This magnetospheric truncation trap has the advantage of stopping the innermost planet in each system approximately at their observed locations, but it is not clear why only some systems contain such short-period planets. Alternate traps further out in the disc and their evolution with time have been considered in some detail by \cite{Hasegawa2011}, showing that different traps move in distinct fashions as the disc evolves. Once a planet is caught in one of these traps, it moves in lock-step with the trap as the disc evolves. This typically occurs on a time-scale much longer time-scale than that of Type I migration, and could allow the planets to survive in the disc until the end of the its lifetime \citep[e.g,][]{Lyra2010,Bitsch2013a}. As mentioned earlier, the larger planets in our simulations may also experience periods of outward migration, depending on their masses and locations at various points during the disc's evolution. \cite{Cossou2014} show that embryos that are large enough to end up in these zones of outward migration generally end up further out as the cores of giant planets, which suggests that the planets in the systems we consider here spent the vast majority of their lifetimes migrating inwards. Overall it is clear that halting the migration of super-Earths and their progenitors is indeed possible, but further work is required to understand these processes in detail.

\subsubsection{Tides}
The proximity of the planets in our models to their stellar hosts suggests that tidal interactions could be a significant contributor to their long-term evolution. Tides can not only circularise planetary orbits, but also shrink them \citep[e.g.,][]{Ogilvie2014}. As a result, it may not be necessary for the innermost planet in each of these systems to have reached its observed position as a result of disc-driven migration. Instead, the planets may have migrated part-way towards their present positions in a disc, with the rest of the migration being a result of tidal dissipation on Gyr time-scales \citep{Hansen2014}. Tidal forces would damp eccentricity in this long phase of evolution, which could in turn affect the results of our stability analysis. As mentioned above, many studies have suggested that tidal forces may also play a role in shaping the distributions of resonances in exoplanetary systems. However, given the uncertainties in the composition of these planets, it is difficult to conduct a more comprehensive study of tidal effects on them at this time.

\section{Summary}
We have performed N-body simulations of the formation of the tightly-packed planetary systems Kepler-11, Kepler-32, and Kepler-80, using parametrized forces to investigate the feasibility of assembling these systems through traditional, disc-driven migration. We find that forming this class of systems via this method is possible under the right circumstances for realistic disc parameters. Our models generally favour strong eccentricity damping and modest levels of disc turbulence, but place no strong constraints on the migration time-scales.  In general we over-predict the incidence of mean-motion resonances, but the significance of this discrepancy is not clear.  We find that disc turbulence cannot explain the low incidence of resonances in compact systems, however, since increased levels of turbulence result in catastrophic disturbances to the systems. Further work is required to understand the paucity of resonances in these systems, and also to investigate the impact of more realistic formation and migration models.

\section*{Acknowledgements}
The authors would like to thank Hossam Aly and Katherine Deck for useful discussions. We also thank the referee, Sean Raymond, for insightful comments which helped to improve this manuscript. TOH is supported by an Science \& Technology Facilities Council (STFC) PhD studentship. RDA acknowledges support from STFC through an Advanced Fellowship (ST/G00711X/1), and from the Leverhulme Trust through a Philip Leverhulme Prize. Astrophysical research at the University of Leicester is supported by an STFC Consolidated Grant (ST/K001000/1). This research used the ALICE High Performance Computing Facility at the University of Leicester. Some resources on ALICE form part of the DiRAC Facility jointly funded by STFC and the Large Facilities Capital Fund of BIS.  This work also used the DiRAC {\it Complexity} system, operated by the University of Leicester IT Services, which forms part of the STFC DiRAC HPC Facility ({\tt http://www.dirac.ac.uk}). This equipment is funded by BIS National E-Infrastructure capital grant ST/K000373/1 and  STFC DiRAC Operations grant ST/K0003259/1. DiRAC is part of the UK National E-Infrastructure.

\bibliographystyle{mn2e}
\bibliography{packed}
\appendix
\section{Damping forces} \label{integrator}
We implement the damping of semi-major-axis and eccentricity via velocity-dependent forces which, when averaged over undamped Kepler orbits, effectuate
\begin{equation}
  \label{eq:damping}
  \dot{a}=-a/\tau_a \quad\text{and}\quad \dot{e}=-e/\tau_e.
\end{equation}
While there are many different possible functional forms for such forces, a simple choice is
\begin{equation} \label{eq:diss}
  \ddot{\B{r}} = -\frac{\D{r}}{2}\frac{1}{\tau_a}
  + \frac{2}{3}\left(\frac{\mu}{h}\,\H{h}\wedge\H{r}-\D{r}\right)\frac{1}{\tau_e}.
\end{equation}
To integrate these forces along with gravity, we extend the kick-drift-kick leapfrog integrator. The latter can be expressed as
\begin{equation}
    K\left(\frac{\Delta t}{2}\right)\;
    D\left(\Delta t\right)\;
    K\left(\frac{\Delta t}{2}\right),
\end{equation}
where $ \Delta t$ is the time step, $K$ the kick operator (which evolves the particles exactly under the gravitational forces for $\D{r}=0$), and $D$ the drift operator (which evolves the particles exactly for $\DD{r}=0$). We implement the damping by extending the leapfrog whilst keeping its time symmetry, hence preserving its second-order accuracy and long-term stability. To this end, we introduce a new operator $E(\Delta t)$ which solves equation~(\ref{eq:diss}) \emph{exactly} for $\D{r}=0$ and write the new integrator as
\begin{equation}
    E\left(\frac{\Delta t}{2}\right)\;
    K\left(\frac{\Delta t}{2}\right)\;
    D\left(\Delta t\right)\;
    K\left(\frac{\Delta t}{2}\right)\;
    E\left(\frac{\Delta t}{2}\right),
\end{equation}
In terms of the radial and azimuthal velocities equation~(\ref{eq:diss}) reads
\begin{eqnarray}
   \dot{v}_r &=& -\left(\frac{1}{2\tau_a}+\frac{2}{3\tau_e}\right) v_r
   \\[0.5ex]
   v_\phi\,\dot{v}_\phi
   &=&\left(\frac{1}{2\tau_a}+\frac{2}{3\tau_e}\right)
      \left[v_e^2-v_\phi^2\right],
\end{eqnarray}
where
\begin{equation}
   v_e^2 = \frac{\mu}{r}\left(1+\frac{3\tau_e}{4\tau_a}\right)^{-1},
\end{equation}
with solution (implementing operator $E$)
\begin{eqnarray}
    v_r(t+ \Delta t) &=& v_r(t)\;\exp\left(-\frac{ \Delta t}{2\tau_a}-\frac{2 \Delta t}{3\tau_e}\right)
        \\[0.5ex]
    v_\phi^2(t+ \Delta t) &=& v_e^2 + \left(v_\phi^2(t)-v_e^2\right)\;
        \exp\left(-\frac{ \Delta t}{\tau_a}-\frac{4 \Delta t}{3\tau_e}\right).
\end{eqnarray}
The resulting damped orbits actually deviate slightly from the ideal~(\ref{eq:damping}), by an amount $\mathcal{O}(T_{\mathrm{orbit}}/ \Delta t)$. However, these deviations are -- to some degree -- physically necessary. For example, semi-major axis damping implies some radial motion and hence eccentricity, even if the orbit was initially circular. In this respect the method differs from that of Lee \& Peale (2002), which simply changes $a$ and $e$ independently, potentially allowing unphysical situations such as a shrinking orbit with $e=0$. We also note that this method can provide considerable efficiency improvements compared to alternatives since the orbital elements do not need to be calculated at each time-step.

\label{lastpage}

\end{document}